\documentclass[aps,prb,reprint,amsmath,amsfsymb]{revtex4-1}
\usepackage{graphicx}
\usepackage{dcolumn}
\usepackage{bm}
\usepackage{color}
\usepackage{siunitx}
\usepackage{caption}
\usepackage{subcaption}
\usepackage{hyperref}
\usepackage{ulem}
\usepackage{todonotes}
\usepackage{threeparttable}
\usepackage{gensymb}
\usepackage{natbib}
\usepackage{url}
\usepackage{textcase}

\begin{document}

\title{Absence of mixed valency for Pr in pristine and hole-doped PrNiO$_2$}
\author{Xingyu Liao}
\affiliation{Department of Physics, University of Illinois at Chicago, Chicago, IL 60607, USA}
\author{Michael R. Norman}
\affiliation{Materials Science Division, Argonne National Laboratory, Lemont, IL, 60439, USA}
\author{Hyowon Park}
\affiliation{Department of Physics, University of Illinois at Chicago, Chicago, IL 60607, USA}
\affiliation{Materials Science Division, Argonne National Laboratory, Lemont, IL, 60439, USA}

\date{\today}

\begin{abstract}
Infinite-layer nickelates ($R$NiO$_2$) exhibit some distinct differences as compared to cuprate superconductors, leading to a debate concerning the role of rare-earth ions ($R$=La, Pr, Nd) in the low-energy many-body physics.
Although rare-earth $4f$ orbitals are typically treated as inert `core' electrons in studies, this approximation has been questioned.  An active participation of $4f$ states is most likely for PrNiO$_2$
based on an analogy to cuprates where Pr cuprates differ significantly from other cuprates.
Here, we adopt density functional plus dynamical mean field theory (DFT+DMFT) to investigate the role of Pr $4f$ orbitals and more generally the correlated electronic structure of PrNiO$_2$ and its hole-doped variant.
We find that the Pr $4f$ states are insulating and show no evidence for either a Kondo resonance or Zhang-Rice singlet formation
as they do not have any hybridization channels near the Fermi energy.
The biggest effects of hole doping 
are to shift the Pr $5d$ and $4f$ states further away from the Fermi energy while enhancing the Ni $3d$ - O $2p$ hybridization, thus reducing correlation effects as the O $2p$ states get closer to the Fermi energy.
We again find no evidence for either Kondo or Zhang-Rice physics for the $4f$ states upon hole doping.
We conclude by commenting on implications for other reduced valence nickelates.

\end{abstract}

\maketitle

\section{Introduction} \label{sec:intro}
Since the infinite-layer nickelate, NdNiO$_2$, was found to be superconducting upon hole doping~\cite{Hwang2019},
intensive research efforts have been made to understand the microscopic origin of the novel properties of these nickelate superconductors. 
Similar to NdNiO$_2$, both PrNiO$_2$ and LaNiO$_2$ are superconducting under hole doping~\cite{LCNO_SC,LSNO_SC,PNO_SC}.
These infinite-layer nickelate superconductors ($R$NiO$_2$) exhibit some similarities and differences with cuprates. 
The crystal structure of $R$NiO$_2$ has Ni$^{1+}$ ions with a $d^9$ electronic configuration in a square-planar geometry, similar to many cuprates.
A major difference compared to cuprates is the weaker hybridization of the Ni $3d_{x^2-y^2}$ states with the O $2p$ states given their larger separation in energy.
This is confirmed by X-ray absorption and resonant inelastic x-ray measurements~\cite{PhysRevB.104.L220505,10.3389/fphy.2021.810220}.
For the same reason, the doped holes in nickelates reside mostly in the Ni $3d_{x^2-y^2}$ orbital, as compared to cuprates where they are mostly O $2p$ in character.
Theoretical studies of $R$NiO$_2$ ($R$=Nd, Pr, and La)~\cite{LNOCCO_theory, NNOCCO_theory} also confirm that the O $2p$ states in these nickelates are located well below the Fermi energy compared to cuprates, leading to infinite-layer nickelates exhibiting Mott-Hubbard-like physics, while the $R$ $5d$ states crossing the Fermi energy mainly produce a self-doping effect.

Although $4f$ states are not relevant for LaNiO$_2$ given that they are unoccupied,
Pr and Nd $4f$ states are partially occupied in $R$NiO$_2$ suggesting the possibility of hybridization with other states near the Fermi energy, though these are typically considered as core electrons in first-principle calculations.
A recent DFT study claimed that Nd $4f$ states modified the Fermi surface of NdNiO$_2$~\cite{NdNiO2_f}, and
a recent many-body calculation argues that these $4f$ states exhibit hybridization with other states resulting in a Kondo resonance at the Fermi energy~\cite{Kang22}.
On the experimental side, a photoemission study of PrNiO$_2$ showed a resonant enhancement of the Pr $4f$ states of PrNiO$_2$ indicating their presence in the valence manifold, though this does not appear to change upon hole doping which would imply weak hybridization of Pr $4f$ with other orbitals~\cite{PNO_exp}.
Moreover, a muon spin rotation experiment suggests a minimal influence of Pr and Nd $4f$ states on the magnetic properties of Ni ions in $R$NiO$_2$~\cite{Fowlie2022}.

On the other hand, Pr ions play a non-trivial role in other reduced valence nickelates as well as cuprates. For example, the ground state of Pr$_4$Ni$_3$O$_8$ is metallic while La$_4$Ni$_3$O$_8$ is a charge-ordered insulator~\cite{Zhang2017}.  
One possibility for this difference is if the Pr $4f$ states hybridize with other states, thus in turn affecting the Ni electronic structure leading to a destabilization of the charge-ordered state.
Moreover, evidence for the hybridization of the Pr $4f$ states with other states has been found in Pr-based cuprates such as PrBa$_2$Cu$_3$O$_7$. In this material, unlike variants with other rare-earth ions, superconductivity is suppressed.
It was argued that the Pr ion is mixed valent due to hybridization with O $2p$ states, giving rise to a Fehrenbacher-Rice $4f$-$2p$ singlet that competes with the $3d$-$2p$ Zhang-Rice singlet~\cite{PhysRevLett.70.3471}.
This is supported by x-ray absorption studies that indicate that Pr has a  higher valence than 3+ in Y$_{1-x}$Pr$_x$Ba$_2$Cu$_3$O$_7$~\cite{YPBCO_mixvalence} and that it deviates even further from 3+ with hole doping by Ca in Pr$_{1-x}$Ca$_x$Ba$_2$Cu$_3$O$_7$~\cite{Staub00}. In the case of PrNiO$_2$ and Pr$_4$Ni$_3$O$_8$, this would give rise to an effective electron doping.
The above observations motivate a further study of the role of $4f$ electrons in $R$NiO$_2$ and its hole-doped variant.

In this paper, we study the correlated electronic structures of LaNiO$_2$ (LNO), PrNiO$_2$ (PNO), La$_{0.75}$Sr$_{0.25}$NiO$_2$ (LSNO), and Pr$_{0.75}$Sr$_{0.25}$NiO$_2$ (PSNO) using DFT+DMFT to investigate the effects of both the Pr $4f$ orbitals as well as hole doping via Sr. 
To construct the LSNO and PSNO structures, we assumed several different supercells as described in Appendix \ref{appdx:struct-rlxns}.  We also did VCA (virtual crystal approximation) calculations as a check (Appendix \ref{appdx:doping-LSNO}).
For DFT+DMFT, we take into account the spin-orbit coupling of the $4f$ orbitals, the large on-site Hubbard interaction of both Pr $4f$ and Ni $3d$ orbitals, and the Hund's coupling. 
A particular focus is to investigate the possible hybridization of Pr $4f$ orbitals as well as how this and the Ni-O hybridization change upon hole doping, and in turn how this affects the correlations of the Ni $3d$ orbitals.

\section{Methods}\label{sec:method}

First, we performed structural relaxations of (La/Pr)NiO$_2$ and their hole-doped variants using the DFT+U method by adopting the Vienna Ab-initio Simulation Package (VASP)~\cite{vasp1,vasp2} (Appendix \ref{appdx:struct-rlxns} for details).
Then, we applied the embedded DMFT method~\cite{eDMFT1, eDMFT2} combined with the full-potential DFT method (WIEN2k)~\cite{WIEN2k_ref} to perform charge self-consistent DFT+DMFT calculations at a temperature $T=300$ K.
We treat Pr $4f$ and Ni $3d$ as correlated orbitals 
that are constructed by using the projector obtained from the DFT bands in the energy window of -10 to 10 eV (relative to the DFT Fermi energy).
For LaNiO$_2$, we did not treat the La $4f$ states as correlated orbitals as they are unoccupied.
For the Hubbard $U$ values, we determined $U=8.3$ eV for the Pr $4f$ orbitals and $U=5.5$ eV for the Ni $3d$ orbitals from the constrained DFT method~\cite{cDFT_ref} (Appendix \ref{appdx:HubbardU}).
We note that the latter is smaller than the value of 7 eV employed in some previous DFT+DMFT studies, though our results do not qualitatively depend on this.
For the Hund's coupling, $J$, we set this to a typical value of 0.6 eV for both Pr $4f$ and Ni $3d$ orbitals, though a larger value of $J$ as used in some previous DFT+DMFT studies is explored in Appendix \ref{appdx:hundJ}.
For the hole-doped variants La$_{0.75}$Sr$_{0.25}$NiO$_2$ (LSNO) and Pr$_{0.75}$Sr$_{0.25}$NiO$_2$ (PSNO),
we used the `mixed' supercell illustrated in Appendix \ref{appdx:struct-rlxns} since it has the most evenly distributed Sr dopant ions among the three supercells we studied, though the results are not sensitive to this choice (Appendix \ref{appdx:struct-rlxns}).
Finally, the many-body occupancies were determined using continuous time quantum Monte Carlo (CTQMC) sampling~\cite{CTQMC_ref}.
The DFT+DMFT spectral functions are obtained using the self-energy data analytically continued to the real frequency axis by adopting the maximum entropy method.

\section{Results}\label{sec:results}

\begin{figure}
    \includegraphics[width=1.1\linewidth]{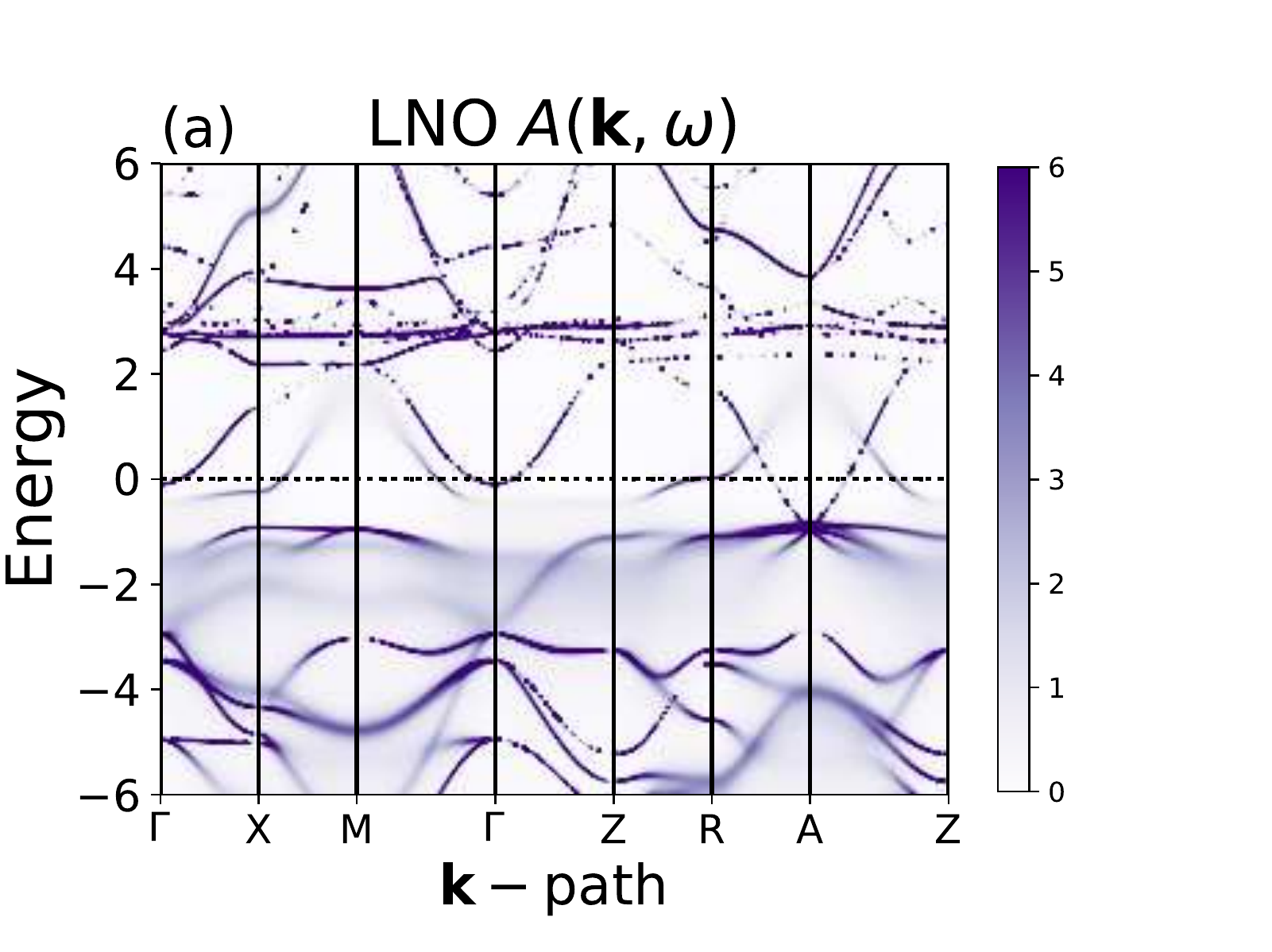}\\
    \vspace{-0.7cm}
    \includegraphics[width=1.1\linewidth]{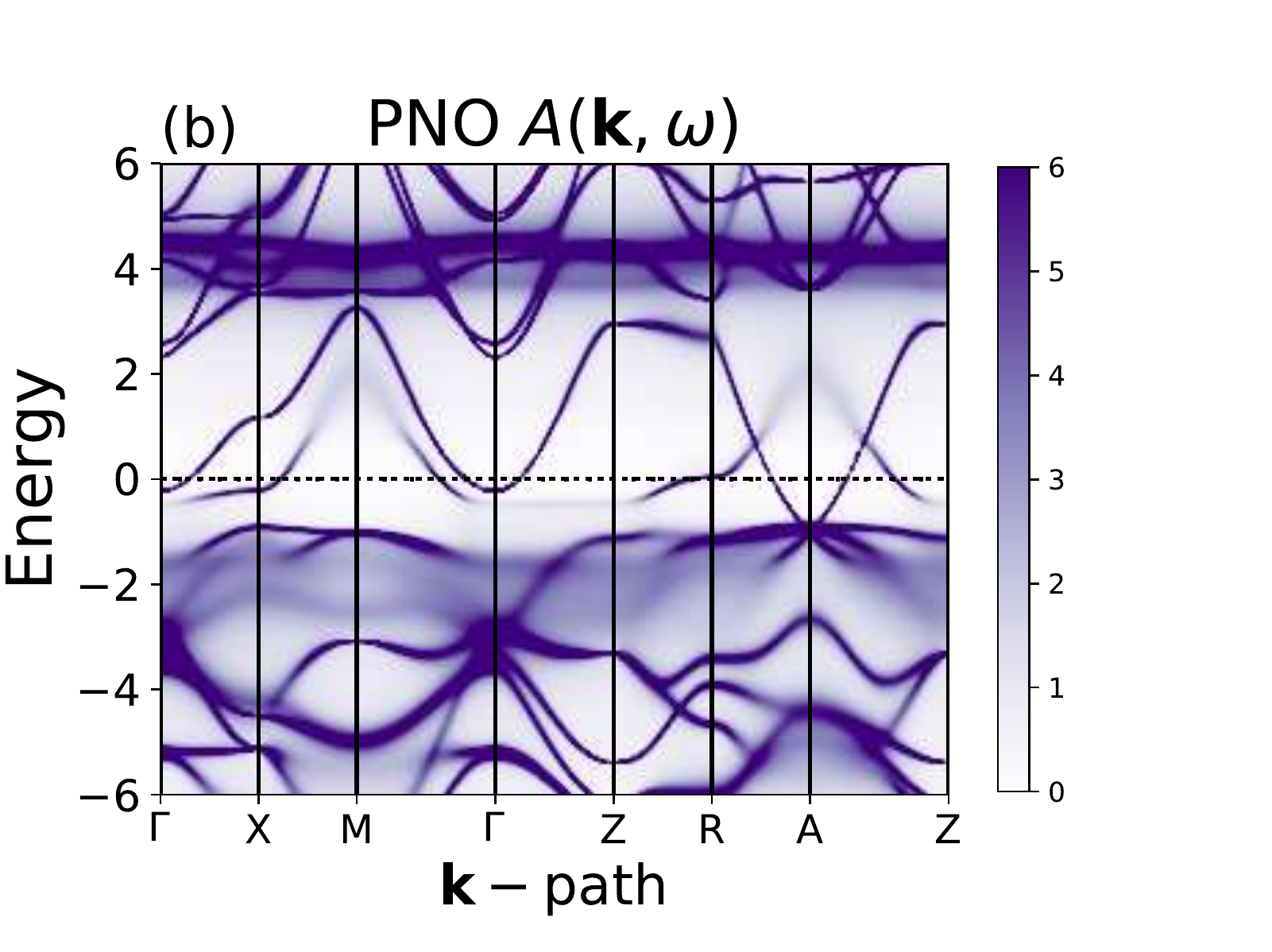}
\caption{The \textbf{k}-resolved spectral function $A(\mathbf{k},\omega)$ obtained from DFT+DMFT for (a) LaNiO$_2$ and (b) PrNiO$_2$.}
\label{fig:BS_DOS_PNO}
\end{figure}

\subsection{Electronic structure: LaNiO$_2$ versus PrNiO$_2$}
In Fig.\:\ref{fig:BS_DOS_PNO}, we plot the \textbf{k}-resolved spectral function $A(\mathbf{k},\omega)$ for LaNiO$_2$ (top panel) and PrNiO$_2$ (bottom panel).
For LaNiO$_2$, the rather dispersionless La $4f$ bands are located about 3 eV above the Fermi energy and are weakly hybridized with the La $5d$ bands.
Our DMFT calculation for PrNiO$_2$ shows several noticeable features compared to LaNiO$_2$.
First, the Pr $4f$ manifold of states splits into two parts, an unoccupied part located about 4 eV above the Fermi energy, and an occupied part about 2 eV below, with this separation due to $U$.  These $4f$ states appear as blurred in the figure
because of the large scattering rate associated with the correlated $4f$ $J=5/2$ orbitals.
The Pr $4f$ $J=7/2$ states, also treated using DMFT, are visible as a thick band at about 4.5 eV above the Fermi energy.
Because all of these states are well away from the Fermi energy, they have a minimal impact on the low
energy physics.  That is, the near $E_F$ electronic structures of LaNiO$_2$ and PrNiO$_2$ are very similar.

We find that the total occupancy of the $4f$ orbitals for PrNiO$_2$ is about 2.07 and as expected, these occupied states (mainly $f^2$ to $f^1$ excitations) have mostly $J=5/2$ character.
While our Pr $4f$ occupancy indicates that the Pr ion is close to the $3+$ valence state, the exact value of the occupancy depends on the choice of the basis functions and the energy window used for its construction. Therefore, we focus on the relative change of the $4f$ (and $3d$) occupancy upon changing doping, etc.
As we demonstrate below from our CTQMC results, we find that Pr is a local $f^2$ $J=4$ ion, consistent
with the $4f$ occupancy we find.
Consistent with this, we find no evidence for a Kondo resonance in contrast
to a recent GW+DMFT study of NdNiO$_2$ that shows such a peak~\cite{Kang22}.

Turning to the other states, the Ni $3d$ bands are mainly located between -3 to -1 eV except $d_{x^2-y^2}$, which is pinned at the Fermi level. 
Most of the O $2p$ spectra are located below -3 eV, although there is some mixture with the Ni $e_g$ bands nearer the Fermi energy. 
The mixing of Ni $e_g$ and O $2p$ states near -1 eV is also evident in the momentum-integrated spectral function shown in Fig.\:\ref{fig:PNO_doping}. 

\subsection{Electronic structure of PrNiO$_2$: Doping effect}
\begin{figure}
    \includegraphics[width=1.02\linewidth]{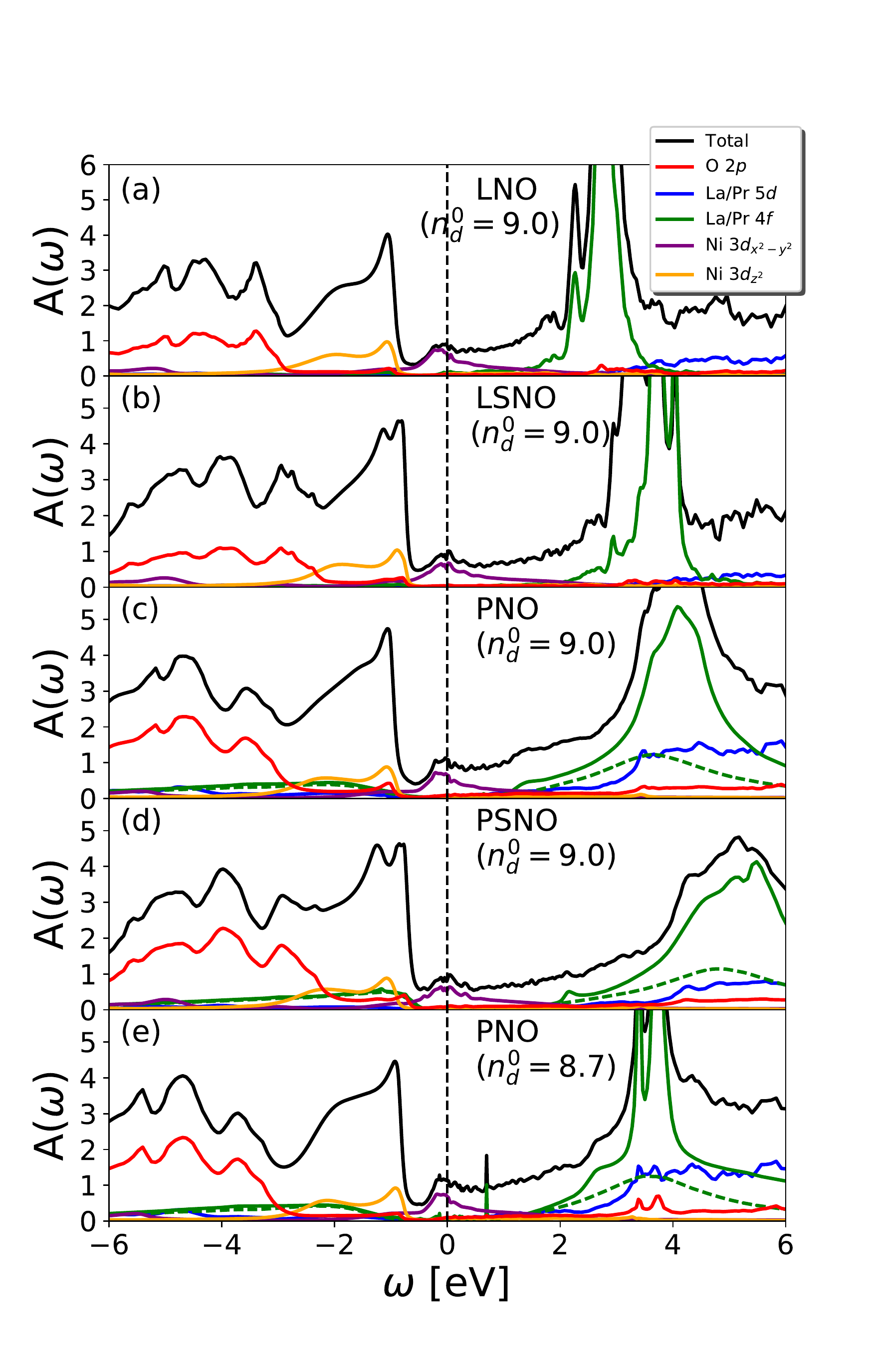}
\caption{The orbital-resolved momentum-integrated spectral function $A(\omega)$ (i.e., the density of states) for (a) LNO, (b) LSNO, (c) PNO, (d) PSNO, and (e) PNO with a different $n_d^0$.
The green dashed curve for the Pr cases shows that part of the $4f$ $A(\omega)$ that has $J=5/2$ character.}
\label{fig:PNO_doping}
\end{figure}

Here, we study the effect of hole doping on the electronic structure as shown in Table \ref{tab:OCC_DC_Prf}.
The 25\% hole doping of PSNO reduces the Ni $d_{x^2-y^2}$ occupancy by 0.07 relative to PNO while the other Ni $3d$ and Pr $4f$ occupancies remain about the same.
Fig.\:\ref{fig:PNO_doping} shows the orbital-resolved
density of states for LNO, LSNO, PNO and PSNO.
Under hole doping, the Pr $4f$ states retain their Hubbard gap, although the unoccupied spectral weight for both Pr $4f$ and $5d$ orbitals shift to higher energy.

This change of the orbital occupancy can be further analyzed from the many-body statistical probabilities shown in Table \ref{tab:weight_DC_Prf}. In all cases, the Ni $d^{10}$ probability 
is below 10\%, consistent with the significantly larger charge transfer energy compared to cuprates.
For the same reason, the Pr-O hybridization is rather weak so that the Pr $f^3$ probability is also less than 10\% (as the $f^1$ probability is very small, there is no tendency for Pr to fluctuate to a 4+ valence state).
That is, Pr is dominated by the $f^2$ configuration, 99.6\% of which is $J=4$ (Table \ref{tab:weight_DC_Prf}). 
The hole doping effect on PSNO decreases the Ni $3d$ occupancy as both Ni $d^9$ and $d^{10}$ probabilities are reduced. 
This hole doping effect also populates the low-spin $S=0$ state more within the Ni $d^8$ sector. Although our DMFT calculation with a Hund's coupling $J=0.6eV$ shows that the high-spin state ($S=1$) is slightly more dominant than the low-spin state ($S=0$) in all cases, a smaller value for $J$ would favor low-spin instead \cite{PhysRevB.103.075123}.  Typically, though, a higher value of $J$ has been assumed in the
nickelate literature, so we show results for $J$=1eV in Appendix \ref{appdx:hundJ}.

In the spectral function $A(\omega)$, for both PSNO and LSNO the Pr/La $5d$ bands are shifted away from the Fermi energy while the occupied O $2p$ bands move closer to the Fermi energy upon hole doping (Fig.$\:$\ref{fig:PNO_doping}). 
The hole doping effect slightly reduces the Pr $4f^3$ probability while again the Pr $4f^1$ probability is almost negligible.
The upshot is that our DFT+DMFT calculation shows that the mixed valence of Pr,
as proposed in Pr-based cuprates~\cite{YPBCO_mixvalence}, is unlikely to happen in these pristine or hole-doped nickelates since the oxygen $2p$ states lie much further away from the Fermi energy.
In more detail, the difference from Ref.~\onlinecite{PhysRevLett.70.3471} is that in their case, the $2p$ levels lie above the $4f$ levels.  Therefore, hybridization pushes the relevant $2p\pi$ levels closer to $E_F$ resulting in their partial occupation (see also Ref.~\onlinecite{PhysRevLett.74.1000}).  Instead, in our case, the $4f$ levels lie {\it above} the $2p$ levels, with the former having their spectral weight away from $E_F$.  As a consequence, hybridization between the $2p\pi$ levels and the relevant $4f$ states, $z(x^2-y^2)$, does not lead to Zhang-Rice singlet formation.

\begin{table}[h]
    \centering
    \caption{Orbital occupancies of LNO, PNO, PSNO, and PNO with a different $n_d^0$ from DFT+DMFT.}
    \begin{tabular}{p{0.18\linewidth}|p{0.16\linewidth}p{0.16\linewidth}p{0.16\linewidth}p{0.16\linewidth}}
    \hline
    Orbitals &         LNO ($n_d^0$=9.0)   &            PNO ($n_d^0$=9.0)  & PSNO ($n_d^0$=9.0) & PNO ($n_d^0$=8.7)  \\
    \hline\hline
    Pr $4f$          & -  &2.07  &2.06  &2.07  \\
    Ni $3d$  & 8.62 & 8.65 & 8.59 & 8.53  \\
    Ni $3d_{x^2-y^2}$  & 1.26 & 1.27 & 1.20 & 1.23  \\
    Ni $3d_{z^2}$  & 1.63 & 1.64     & 1.63 & 1.58  \\
    \hline
    \end{tabular}
    \label{tab:OCC_DC_Prf}
\end{table}

\begin{table}[h]
    \centering
    \caption{Statistical probabilities (in \%) of the many-body states of LNO, PNO, PSNO, and PNO with a different $n_d^0$ from DFT+DMFT.}
    \begin{tabular}{p{0.17\linewidth}|p{0.16\linewidth}p{0.16\linewidth}p{0.16\linewidth}p{0.16\linewidth}}
    \hline
    States &         LNO ($n_d^0$=9.0)  &   PNO ($n_d^0$=9.0)  & PSNO ($n_d^0$=9.0) & PNO ($n_d^0$=8.7)  \\
    \hline\hline
    Pr $f^2$          & -   & 92.0  & 93.0  & 92.0 \\
    Pr $f^2_{J=4}$ & - & 91.6 & 92.5 & 91.6 \\
    Pr $f^3$          & -   & 7.2  & 6.1  & 7.2 \\
    Ni $d^8_{S=0}$  & 8.9 & 12.2 & 17.0 & 14.0  \\
    Ni $d^8_{S=1}$  & 26.2 & 21.4   & 19.5 & 25.3 \\
    Ni $d^9$  & 51.0 & 52.8     & 50.7 & 47.8 \\
    Ni $d^{10}$  & 8.3 &  8.7  & 7.2 & 6.1 \\
    \hline
    \end{tabular}
    \label{tab:weight_DC_Prf}
\end{table}

\begin{figure}
    \includegraphics[width=1.0\linewidth]{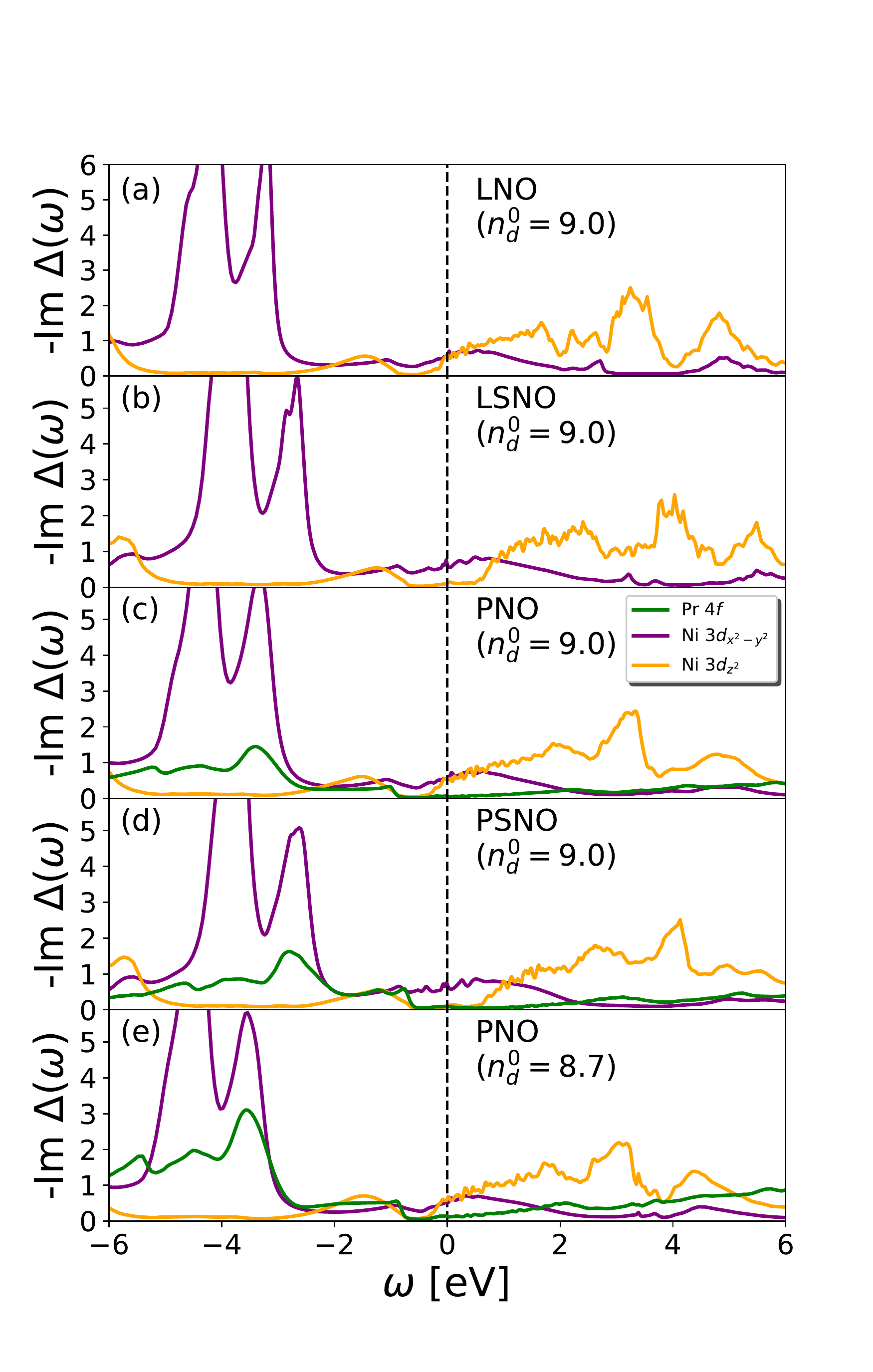}
\caption{The hybridization function $\Delta(\omega)$ for (a) LNO, (b) LSNO, (c) PNO, (d) PSNO, and (e) PNO with a different $n_d^0$.}
\label{fig:PNO_delta}
\end{figure}

\begin{figure}
    \includegraphics[width=1.0\linewidth]{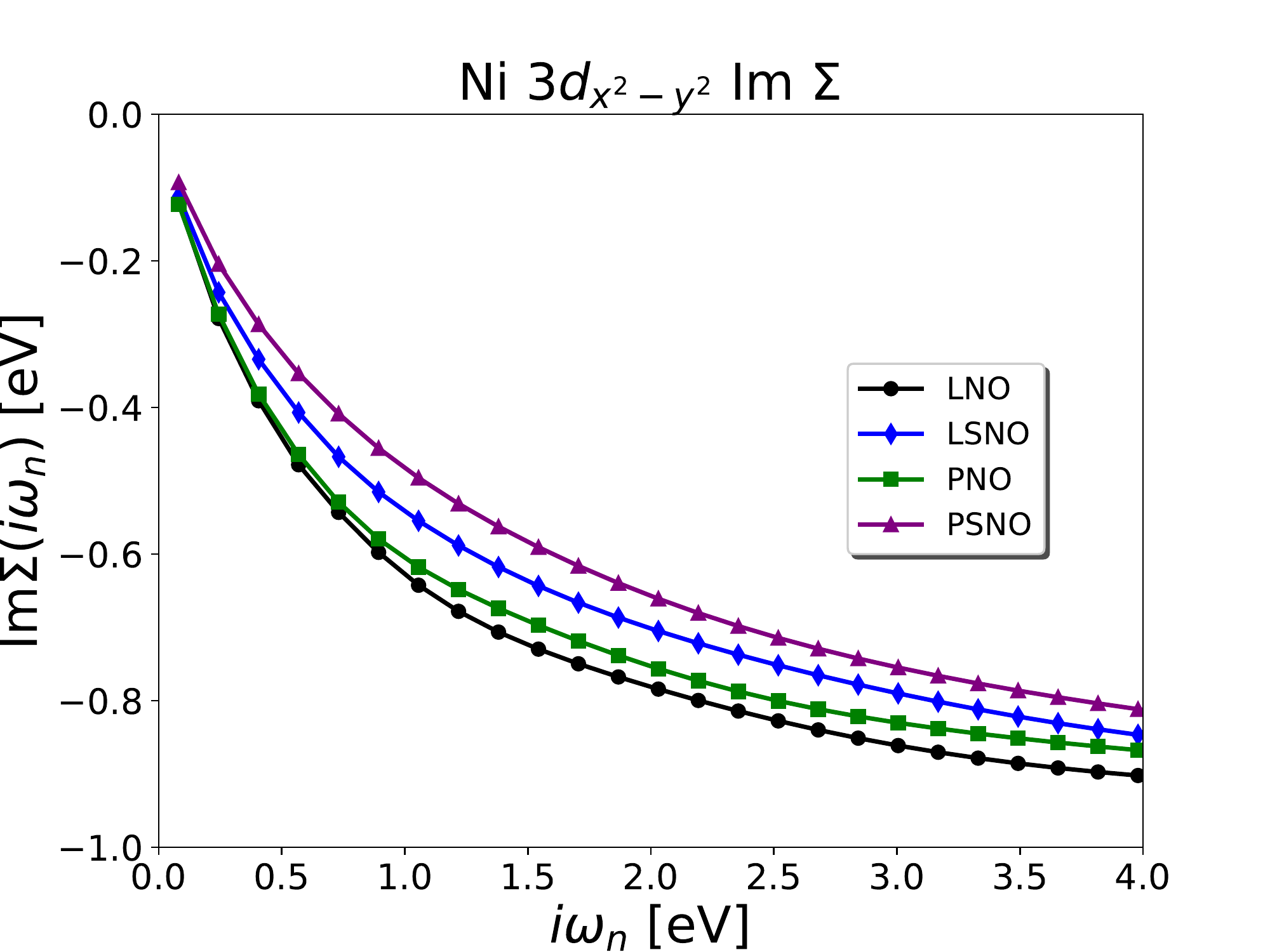}
\caption{The imaginary part of the self energy $\Sigma(i\omega_n)$ for Ni $3d_{x^2-y^2}$ orbitals in LNO, LSNO, PNO, and PSNO with $n_d^0=9.0$.}
\label{fig:PNO_sigma1}
\end{figure}

Next, we address the correlations of the Ni $d_{x^2-y^2}$ orbital as it is the most relevant near $E_F$.
We compute the mass renormalization, $m^*/m$, from the self-energy on the Matsubara frequency axis $\omega_n$:
\begin{equation}
    \frac{m^*}{m}=1-\frac{\partial Im\Sigma(i\omega_n )}{\partial \omega_n}\vert_{\omega_n\rightarrow 0}
\end{equation}
where $Im\Sigma$ is the imaginary part of the self-energy.
Here, we fit $Im\Sigma$ (Fig.\:\ref{fig:PNO_sigma1}) to a polynomial of the fourth-order in $\omega_n$ using the lowest six Matsubara points as in previous DMFT calculations~\cite{PhysRevB.102.245130}.
In Table \ref{tbl:mass}, we list $m^*/m$ for LNO, LSNO, PNO, and PSNO.
Although there is no significant difference of $m^*/m$ between LNO and PNO, the hole doping significantly reduces $m^*/m$ for LSNO and PSNO as the correlation effects weaken. 
This change of $m^*/m$ can be understood from the hybridization function shown in Fig.\:\ref{fig:PNO_delta}.
With hole doping, the oxygen states move closer to $E_F$ for both LSNO and PSNO.
At the same time, the probability of Ni $d^9$ states is also reduced while the Ni $d^8$ states are more populated.
This demonstrates that upon hole doping, most holes reside on the Ni sites as expected in the Mott limit
in contrast with cuprates where the doped holes are mainly on the oxygen sites.

For the Ni $3d_{z^2}$ orbital, $m^*/m$ does not change much for the different cases as the Ni $3d_{z^2}$ occupancy is nearly the same and the Ni $3d_{z^2}$ - O $2p$ hybridization is also weaker than that for $d_{x^2-y^2}$ (Fig.$\:$\ref{fig:PNO_delta}).
Interestingly, the unoccupied $3d_{z^2}$ states exhibit hybridization with the Pr $5d$ states over a broad range in energies in all cases, which was also inferred from a recent RIXS measurement on La$_4$Ni$_3$O$_8$~\cite{Yao23}.

\begin{table}[h]
    \centering
    \caption{The mass renormalization of LNO, PNO, PSNO, and PNO with a different $n_d^0$ from DFT+DMFT.}
    \begin{tabular}{p{0.18\linewidth}|p{0.14\linewidth}p{0.14\linewidth}p{0.14\linewidth}p{0.14\linewidth}p{0.14\linewidth}}
    \hline
    Orbitals &         LNO ($n_d^0$=9.0)  &    LSNO ($n_d^0$=9.0) &             PNO ($n_d^0$=9.0)  & PSNO ($n_d^0$=9.0) & PNO ($n_d^0$=8.7)  \\
    \hline\hline
    Ni $3d_{x^2-y^2}$  & 1.98 & 1.75 &1.94 & 1.68 & 2.06           \\
    \hline
    Ni $3d_{z^2}$  & 1.26 & 1.25 & 1.26 & 1.25 & 1.28           \\
    \hline
    \end{tabular}
    \label{tbl:mass}
\end{table}

\subsection{Double counting effect on Ni $3d$ orbitals}
Here, we explore the effect of the double counting potential on the electronic structures and occupancies of Ni $3d$ states in PNO within DFT+DMFT. 
The double counting potential $V_{DC}$ for the Ni $3d$ orbitals can be calculated using the following formula~\cite{PhysRevB.90.075136,dmft_nominaldc}:
\begin{equation}
\label{eq:double_counting_nominal}
    V_{DC}=U(n_d^0-\frac{1}{2})-\frac{J}{2}(n_d^0-1)
\end{equation}
where $n_d^0$ is typically chosen as the nominal occupancy of the $3d$ orbitals.
Although $n_d^0$ is an integer in the atomic limit, it is not clear what $n_d^0$ would be appropriate due to hybridization.
For PrNiO$_2$, Ni $3d$ is self-doped due to the partially occupied Pr $5d$ bands. As a result, the $3d$ occupancy is close to 8.7 in contrast with the nominal occupancy in the atomic limit of 9.0 while the $4f$ occupancy is close to the integer value ($n_f^0=2.0$).
Although a previous DMFT calculation of the effect of hole doping for NdNiO$_2$ was performed using $V_{DC}$ with $n_d^0=$9.0~\cite{NdNiO2_doping}, we explore here the effect of a different $V_{DC}$ by using $n_d^0=$8.7 as shown in Tables \ref{tab:OCC_DC_Prf}, \ref{tab:weight_DC_Prf},
\ref{tbl:mass} and Figs.~\ref{fig:PNO_doping}, \ref{fig:PNO_delta}.
This reduced $V_{DC}$ decreases the Ni-O hybridization while the other non-correlated orbitals are not affected, and as a result, the $d^{10}$ state is less populated while the $d^8$ state is more populated as shown in Table \ref{tab:weight_DC_Prf}.
This reduced hybridization also gives rise to a larger mass renormalization of the Ni $3d_{x^2-y^2}$ orbital (Table \ref{tbl:mass}), meaning that the correlation physics of the Ni $3d_{x^2-y^2}$ orbital is mostly dictated by the Ni-O hybridization.
However, this reduced $V_{DC}$ does not affect the correlation effect on Pr $4f$ $J=5/2$ states as shown in Fig.$\:$\ref{fig:PNO_doping} (green dashed curves) as their self-energy changes are not noticeable (see Fig.\:\ref{fig:PNO_sigma2} and Fig.\:\ref{fig:PNO_sigma3}). 
Nevertheless, the spectral function for the entire Pr $4f$ manifold of states (Fig.$\:$\ref{fig:PNO_doping}, green solid curves) exhibits more sharp features in energy for PNO with $n_d^0=8.7$ as the Pr $4f$ $J=7/2$ states are less affected by the scattering effects from the DMFT self-energy.

We have also explored the effect of varying $V_{DC}$ for the $4f$ orbitals.
We tested three different $n_f^0$ values ($n_f^0$= 2.0, 1.8, and 1.5) and found that the Pr $f^2$ state is always dominant ($\sim$92-94\%), the resulting 4$f$ occupancies are always close to 2.0, and the spectral gap for the $4f$ states is always present.

\subsection{The Mott transition of Pr $4f$ states}

\begin{figure}
    \includegraphics[width=1.0\linewidth]{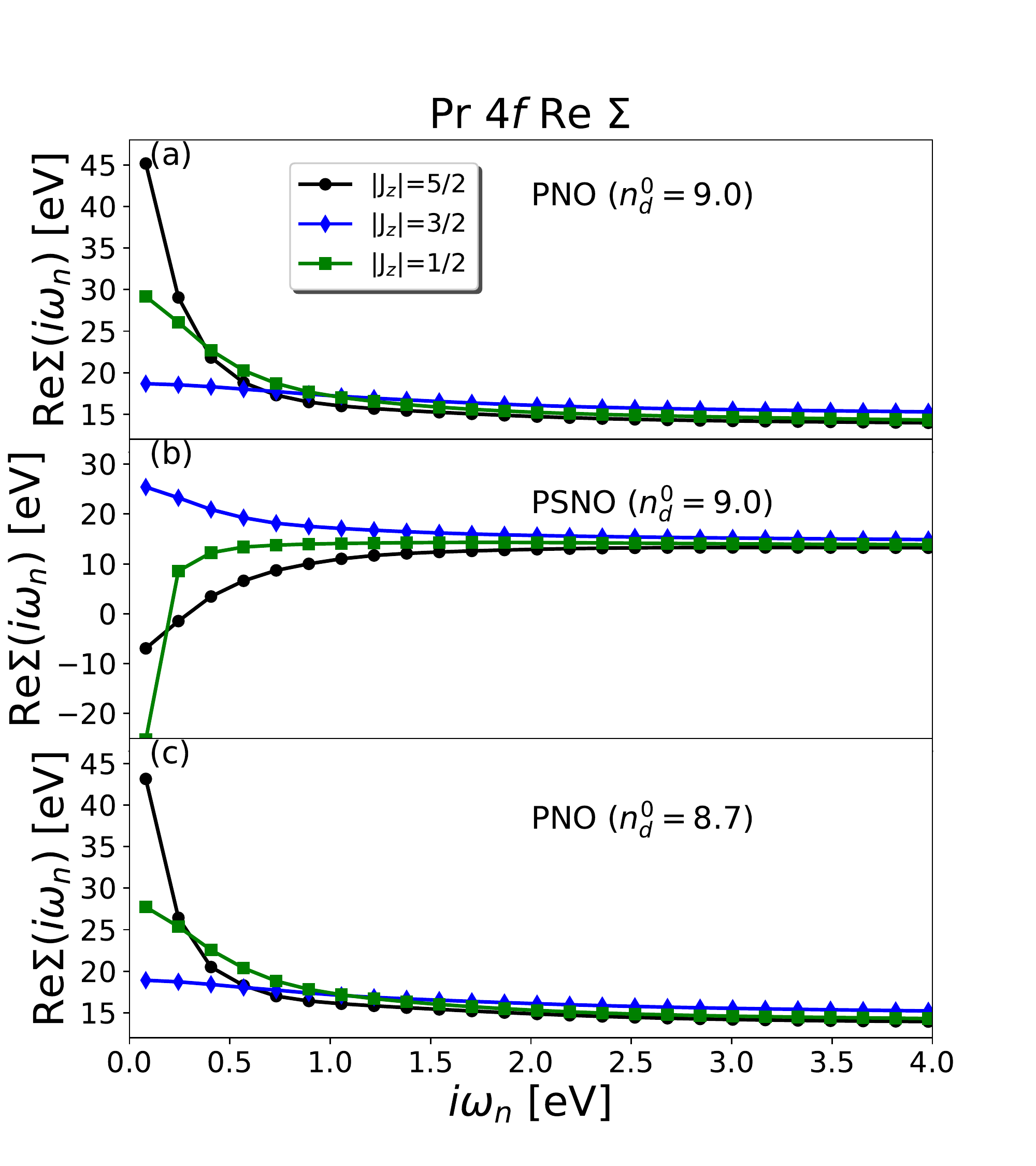}
\caption{The real part of the self energy $\Sigma(i\omega_n)$ for Pr $4f$ orbitals in (a) PNO with $n_d^0=9.0$, (b) PSNO with $n_d^0=9.0$, and (c) PNO with $n_d^0=8.7$.}
\label{fig:PNO_sigma2}
\end{figure}

\begin{figure}
    \includegraphics[width=1.0\linewidth]{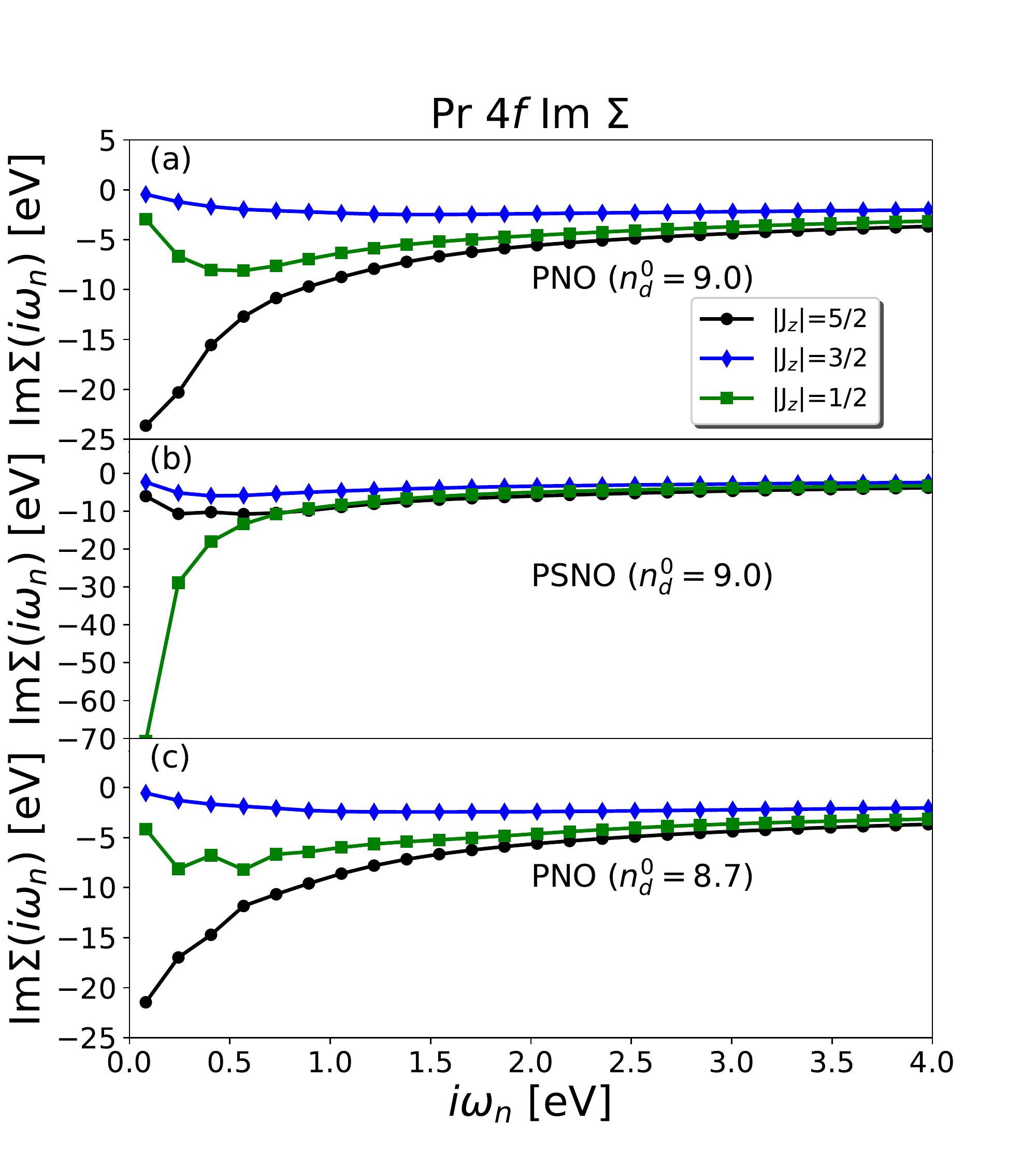}
\caption{The imaginary part of the self energy $\Sigma(i\omega_n)$ for Pr $4f$ orbitals in (a) PNO with $n_d^0=9.0$, (b) PSNO with $n_d^0=9.0$, and (c) PNO with $n_d^0=8.7$.}
\label{fig:PNO_sigma3}
\end{figure}

\begin{figure}
    \includegraphics[width=1.0\linewidth]{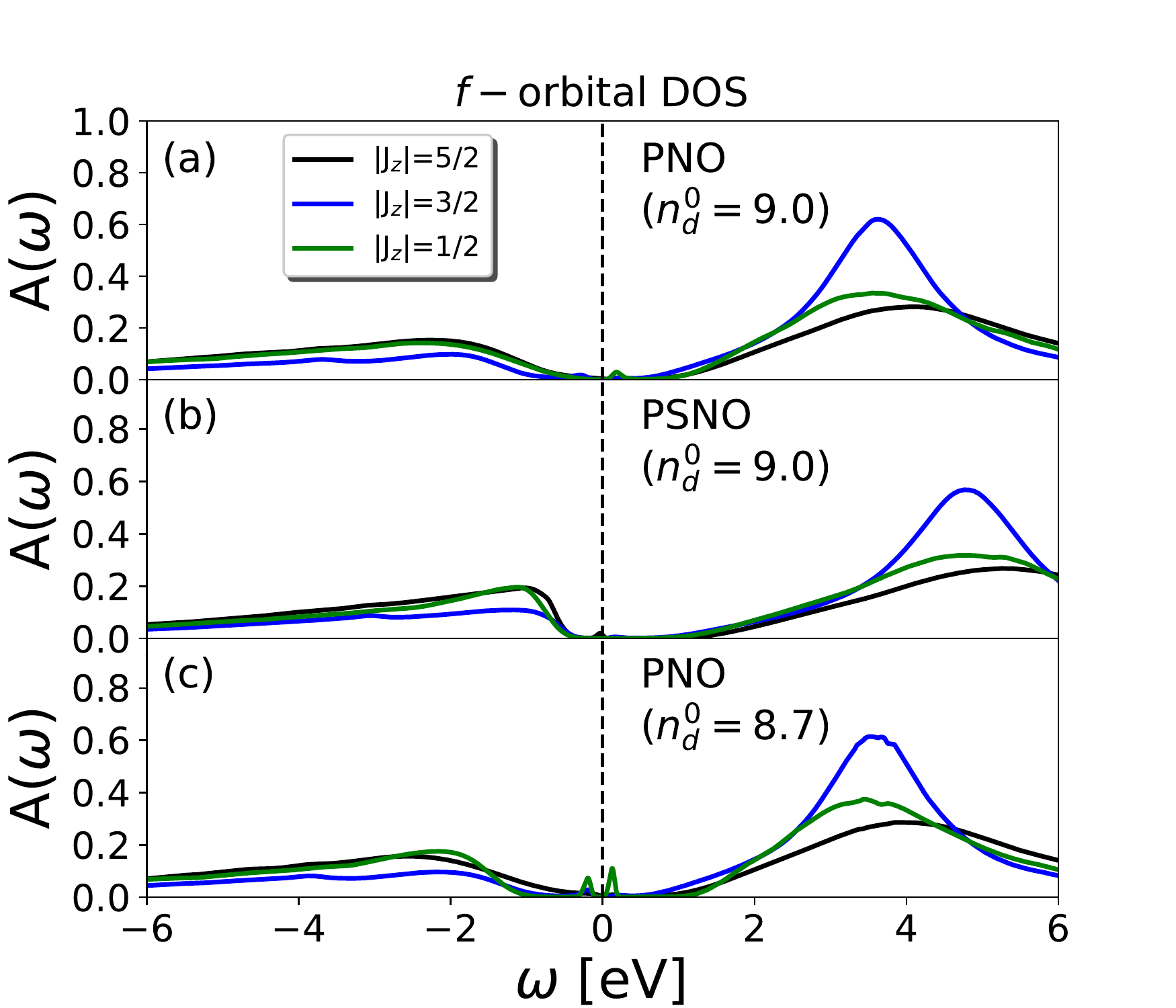}
\caption{The momentum-integrated spectral function $A(\omega)$ for Pr $4f$ $J=5/2$ orbitals in (a) PNO with $n_d^0=9.0$, (b) PSNO with $n_d^0=9.0$, and (c) PNO with $n_d^0=8.7$.}
\label{fig:PNO_f}
\end{figure}

\begin{figure}
    \includegraphics[width=1.0\linewidth]{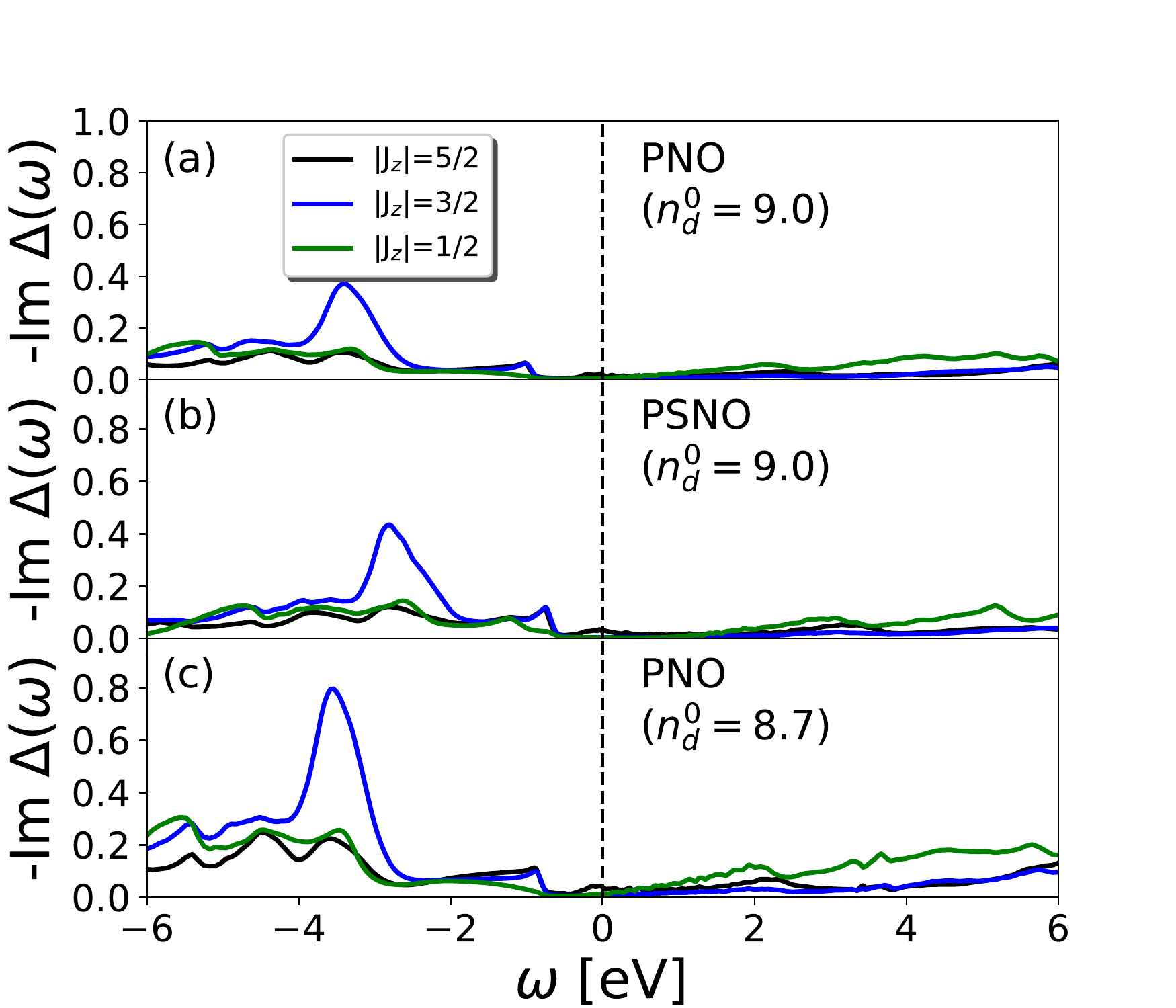}
\caption{The hybridization function for Pr $4f$ $J=5/2$ orbitals in (a) PNO with $n_d^0=9.0$, (b) PSNO with $n_d^0=9.0$, and (c) PNO with $n_d^0=8.7$.}
\label{fig:PNO_f2}
\end{figure}

We find insulating behavior for the Pr $4f$ states given the rather large $U$ of 8.3 eV that we obtain from our constrained DFT method (Appendix \ref{appdx:HubbardU}) along with the lack of hybridization channels near $E_F$ (Fig.~\ref{fig:PNO_f2}).
To study the nature of these insulating states, we plot both the real and imaginary parts of the $J=5/2$ Matsubara self energies in Fig.\:\ref{fig:PNO_sigma2} and Fig.\:\ref{fig:PNO_sigma3},
noting that the occupied part of $A(\omega)$ represents mainly $f^2$ to $f^1$ excitations and
the unoccupied part mainly $f^2$ to $f^3$ excitations.
Here, we decompose this in a $J_z$ basis, with results presented in a cubic basis in Appendix \ref{appdx:hundJ} (Pr being surrounded by a slightly squashed cube of oxygen ions).
As expected from this oxygen environment around the Pr,
the $|J_z|=3/2$ states (dominantly $\Gamma_7$ character) show a larger unoccupied spectral weight compared to the other $J_z$ states (Fig.\:\ref{fig:PNO_f}).
Moreover, the occupied $|J_z|=3/2$ states are more hybridized with O $2p$ orbitals below $\sim$ -3eV from the Fermi energy compared to the other $J_z$ states (Fig.\:\ref{fig:PNO_f2}), as expected since these states have a large $z(x^2-y^2)$ component that points towards the oxygen sites (this is the hybridization central to the $4f-2p$ singlet formation in  Refs.~\onlinecite{PhysRevLett.70.3471,PhysRevLett.74.1000}).
The $|J_z|=1/2$ and $5/2$ states are approximately degenerate as they are mostly $\Gamma_8$ in character.
Indeed, in PNO ($n_d^0=9.0$), both the real and imaginary parts of the low-energy self-energies for $|J_z|=1/2$ and $5/2$ states show a pole-like structure, $1/(i\omega_n-\epsilon)$, where $\epsilon$ is the position of the pole (see Fig.\:\ref{fig:PNO_sigma2} and Fig.\:\ref{fig:PNO_sigma3}), while the self-energies for $|J_z|=3/2$ do not exhibit this behavior.
This pole-like behavior is most visible for the $|J_z|=5/2$ states that have a more typical Mott insulating behavior where a diverging $Im \Sigma$ as $\omega\rightarrow0$ is seen since the pole of the self-energy is close to the Fermi energy.

Under hole doping, both $|J_z|=5/2$ and $1/2$ states still exhibit Mott insulating-like behavior, although it is now $Im\Sigma$ for the $|J_z|=1/2$ states that diverges as $\omega\rightarrow0$ since the position of its self-energy pole has shifted closer to the Fermi energy. Therefore, the position of these self-energy poles are sensitive to the hole doping due to the shift in chemical potential.
The change of the double-counting effect for the $3d$ electrons ($n_d^0$) does not alter much either the self-energy or the spectral function behaviors for the Pr $4f$ $J=5/2$ states. 


\section{Summary}
We studied the correlated electronic structure of Pr $4f$ and Ni $3d$ orbitals for both pristine and hole-doped PrNiO$_2$ using DFT+DMFT.
We find that the Pr $4f$ states in PrNiO$_2$ do not affect the low-energy physics of the Ni $3d$ orbitals, so it is similar to LaNiO$_2$ where the $4f$ states are unoccupied. These Pr $4f$ states are gapped both due to $U$ and the lack of any hybridization channels with other orbitals near the Fermi energy, however this spectral gap is smaller than $U$ since the Pr $4f$ states still hybridize weakly with Pr $5d$ and O $2p$ states at the relevant energies, the latter indicated by the finite $f^3$ weight shown in Table \ref{tab:weight_DC_Prf}.
Regardless, Pr behaves like a local $f^2$ $J=4$ ion.

The hole-doping effect of replacing Pr by Sr in PrNiO$_2$ mainly shifts the Pr $4f$ and $5d$ states in energy (due to the shift in chemical potential) and also reduces the occupancy of the Ni $3d_{x^2-y^2}$ states.
As a result, the many-body configurations of Pr $f^3$ and Ni $d^{10}$ are less populated (Table \ref{tab:weight_DC_Prf}) along with a smaller possibility for mixed valency as the spectral weight of the unoccupied Pr $5d$ and O $2p$ states becomes even weaker as they move further away from the Fermi energy.
However, this hole-doping effect also reduces the mass renormalization of the Ni $3d_{x^2-y^2}$ orbital as the occupied O $2p$ states move closer to the Fermi energy, and it enhances the Ni $d^8_{S_z=0}$ (low-spin) probability by populating more Ni $d^8$ states, thus taking the material further away from the Mott limit where $d^9$ dominates.
We also find that the correlation effect on the Ni $d_{x^2-y^2}$ orbital is mostly dictated by the Ni-O hybridization, as the mass renormalization of the Ni $3d_{x^2-y^2}$ orbital is increased by a reduced Ni-O hybridization as demonstrated for PNO by reducing the double counting potential.

In conclusion, we find no evidence for mixed valency for Pr in PrNiO$_2$ or its hole-doped variant, in contrast to what has been suggested in Pr-cuprates.  This seems to be consistent with available data for infinite-layer nickelates.  This would also argue that the difference between La$_4$Ni$_3$O$_8$ and Pr$_4$Ni$_3$O$_8$ mentioned at the beginning of this paper is unlikely due to mixed valency.  This would be consistent with x-ray absorption data at the Pr M$_4$ and M$_5$ edges that found no evidence of mixed valency~\cite{Zhang2017}.  Still, a small valence change in Pr due to the smaller charge transfer energy of Pr$_4$Ni$_3$O$_8$ compared to PNO and PSNO could conceivably destabilize the charge order insulator found in La$_4$Ni$_3$O$_8$.  This could be investigated further by a similar DFT+DMFT study as done in this paper.

\section*{Acknowledgement}
We thank Kristjan Haule for discussions concerning our DMFT results.
XL was supported by ACS-PRF grant 60617. HP and MN were supported by the Materials Sciences and Engineering Division, Basic Energy Sciences, Office of Science, US Dept.~of Energy. We gratefully acknowledge the computing resources provided on Bebop, a high-performance computing cluster operated by the Laboratory Computing Resource Center at Argonne National Laboratory.

\bibliography{main.bib}

\clearpage

\appendix
\section{Calculation of Hubbard $U$ parameters}
\label{appdx:HubbardU}
\begin{figure}
    \centering
    \includegraphics[width=1\linewidth]{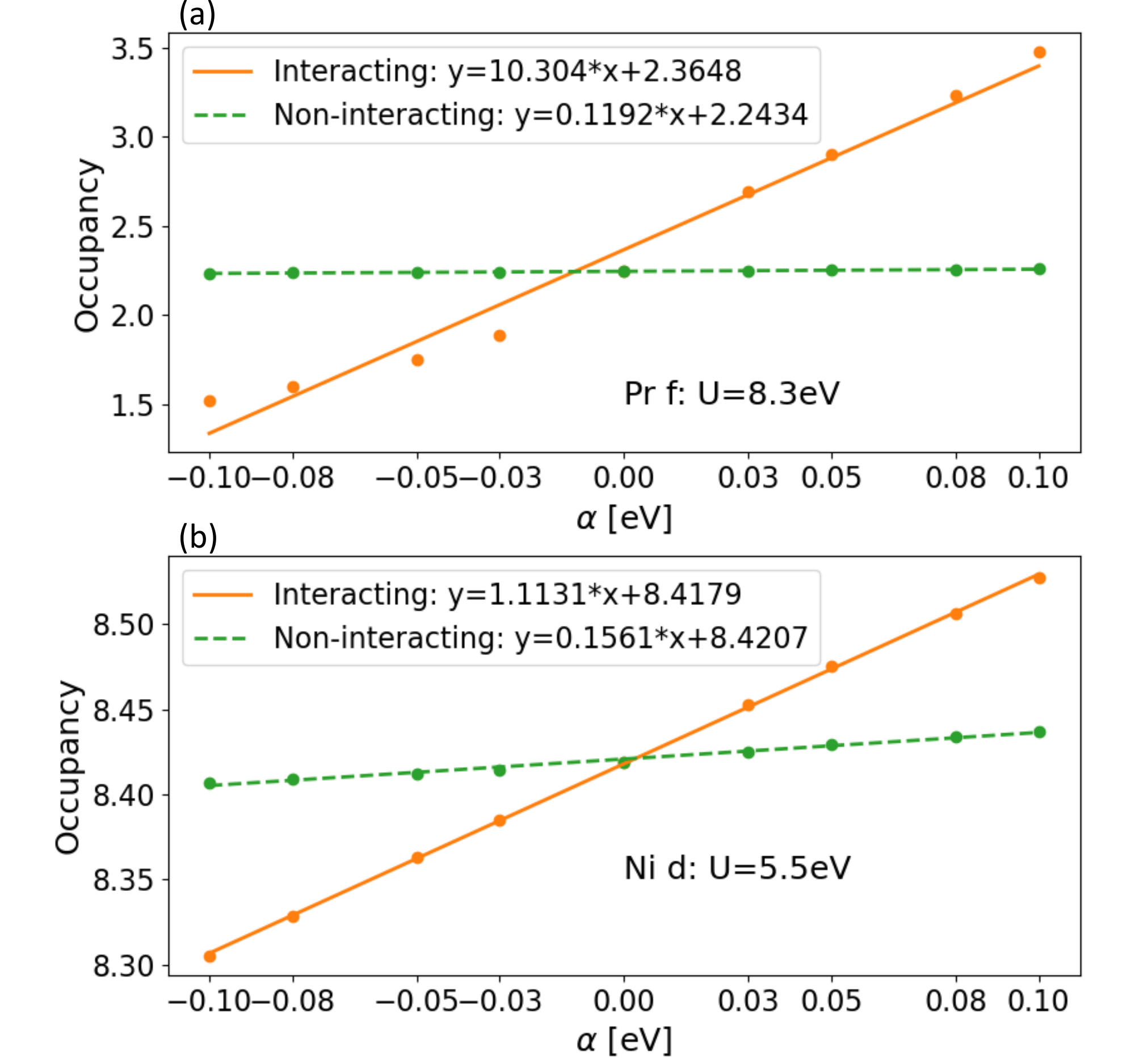}
\caption{Constrained DFT occupancy change as a function of an applied local potential, $\alpha$, for (a) Pr $4f$ and (b) Ni $3d$ orbitals. The fitted slopes and calculated Hubbard $U$ values are listed on the plots.}
\label{fig:cDFT_PrNi}
\end{figure}

Here, we compute the Hubbard $U$ values for Ni $3d$ and Pr $4f$ orbitals in (Pr/La)NiO$_2$
by adopting the constrained DFT (cDFT) method based on first-principles. 
We use the same $U$ values for their hole-doped variants.
Following the scheme of cDFT, $U$ of the site \textit{I} can be calculated from
\begin{equation}
    \label{eq:cdft_U}
    U=(\chi^{-1}_0-\chi^{-1})_{II}
\end{equation}
Here, $\chi^{-1}_0$ and $\chi^{-1}$ are the non-interacting and interacting density response functions of the system related to a local perturbation $\alpha$:
\begin{eqnarray}
    \label{eq:cdft_LR}
    &\chi^0_{IJ}=\frac{\partial{n_I^{KS}}}{\partial\alpha_J}\\
    &\chi_{IJ}=\frac{\partial{n_I}}{\partial\alpha_J}
\end{eqnarray}
where $\alpha_J$ is a localized applied potential at site \textit{J} and $n_I$ ($n_I^{KS}$) is the occupancy at site \textit{I} from an interacting (non-interacting) calculation ($KS$ denotes the Kohn-Sham bands).
In practice, the interacting calculation refers to the regular charge self-consistent calculation and the non-interacting calculation means that the charge density is fixed during the calculation without the feedback of the self-consistent interaction effect.
Here, we adopt the Vienna Ab-initio Simulation Package (VASP)~\cite{vasp1,vasp2} for cDFT calculations.
Occupancies of correlated orbitals obtained at different potential $\alpha$ values are shown in Table \ref{tbl:cDFT_occ} and the plot of the occupancy as a function of $\alpha$ from which $U$ is obtained is shown in Fig.~\ref{fig:cDFT_PrNi}. 
The slope of the interacting (non-interacting) line is $\chi^{-1}$ ($\chi^{-1}_0$).
The calculated $U$ value for the Ni $3d$ orbitals is 5.5 eV, which is close to
the experimental estimate of 5 eV for PrNiO$_2$~\cite{PNO_exp}.
For Nd and Ce, the $U$ value of $4f$ orbitals in previous theoretical work was set to 6 eV~\cite{Ce_fU_dmft, NdNiO2_doping}. 
In our calculation, we find $U$ for the Pr $4f$ orbitals to be 8.3 eV. 

\begin{table}
    \centering
    \caption{Occupancy obtained from different local potentials, $\alpha$, for Ni $3d$ and Pr $4f$ orbitals. Here, $\alpha$ is in eV units, `int' denotes interacting, and `non-int' non-interacting.} 
    \begin{tabular}{p{0.1\linewidth}|p{0.2\linewidth}p{0.15\linewidth}|p{0.2\linewidth}p{0.15\linewidth}}
    \hline\hline
        &\multicolumn{2}{c}{Ni 3d}    &      \multicolumn{2}{c}{Pr 4f} \\
        \hline
      $\alpha$ & int & non-int & int & non-int\\ 
    \hline
    -0.1    &8.305  &8.407    &1.517	&2.231\\
    -0.08   &8.328	&8.409    &1.598	&2.234\\
    -0.05   &8.363	&8.412    &1.747	&2.238\\
    -0.03   &8.385	&8.414    &1.885	&2.240\\
    0       &8.419	&8.419    &2.244	&2.244\\
    0.03    &8.453	&8.425    &2.689	&2.246\\
    0.05    &8.475	&8.429    &2.899	&2.250\\
    0.08    &8.506	&8.434    &3.228	&2.252\\
    0.1     &8.527	&8.437    &3.476	&2.256\\
    \hline
    \end{tabular}
\label{tbl:cDFT_occ}
\end{table} 

\section{Structural relaxations}
\label{appdx:struct-rlxns}

In this study, we performed structural relaxations of (La/Pr)NiO$_2$ and their hole-doped variants. 
We adopt the Vienna Ab-initio Simulation Package (VASP)~\cite{vasp1,vasp2} to perform the structural relaxations using the DFT+U method.
The Perdew-Burke-Ernzerhof for solids (PBE-sol)~\cite{PBEsol} exchange and correlation energy functional was adopted. 
We used a 10$\times$10$\times$10 $k$ point grid and converged the atomic forces of all ions to be smaller than 0.001 eV/\AA $\:$ for all structural relaxations of the cell shape, volume and internal ionic positions.
We used DFT+U with $U$ values obtained from the cDFT method (Appendix \ref{appdx:HubbardU}), namely $U$=5.5eV for Ni $3d$ and $U$=8.3eV for Pr $4f$ orbitals. We didn't apply $U$ values to La $4f$ orbitals and used the Hund's coupling $J$=0.6eV for all structures, to be consistent with $U$ and $J$ values used in the main text. For DFT+U, we used the ferromagnetic spin configuration.

\begin{figure}
    \centering
    \includegraphics[width=0.95\linewidth]{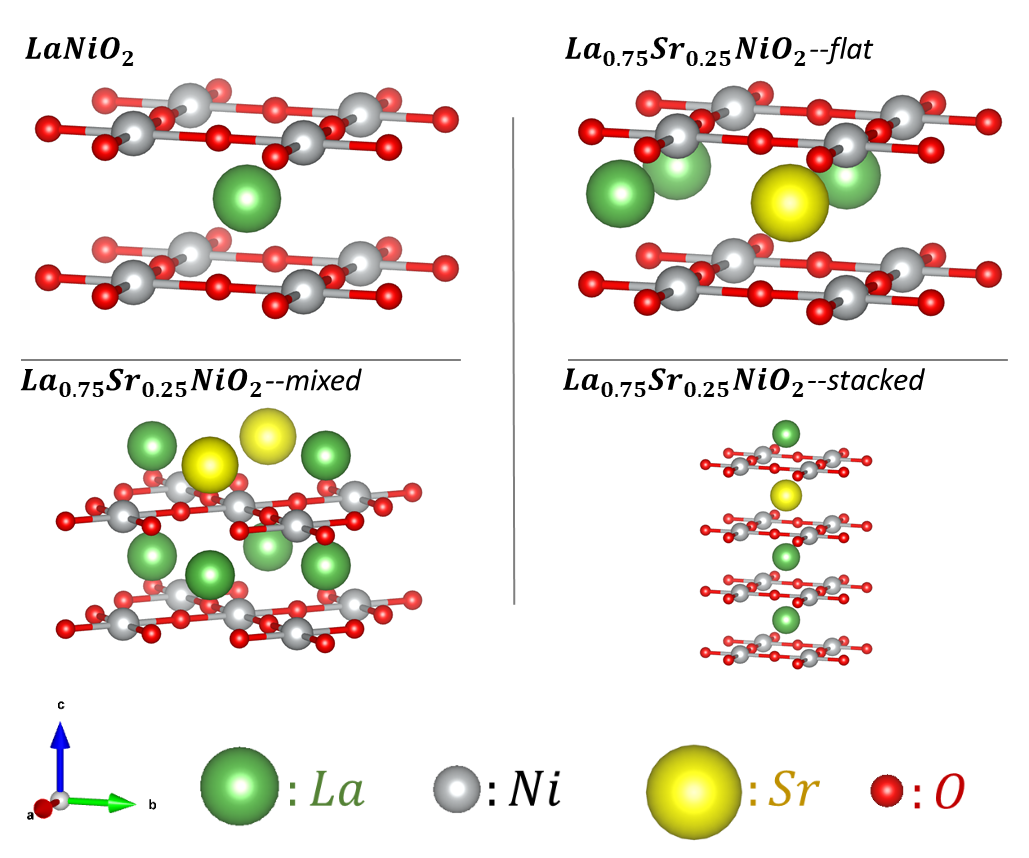}
\caption{Structures of LaNiO$_2$ (LNO) and La$_{0.75}$Sr$_{0.25}$NiO$_2$ (LSNO). Three different supercells were considered for LSNO: a flat structure (upper right panel), a stacked structure (lower right panel) and a mixed structure (lower left panel).}
\label{fig:LSNO_struc}
\end{figure}

First, we relaxed pristine (La/Pr)NiO$_2$ structures in the tetragonal unit cell ( Fig.~\ref{fig:LSNO_struc}). Due to the smaller size of Pr compared to La, the $a(c)-$axis of PNO is 3.88(3.32)\AA\:, which is slightly reduced by 1-1.5\% compared to the LNO $a(c)-$axis of 3.94(3.35)\AA\:.
For La$_{0.75}$Sr$_{0.25}$NiO$_2$ (LSNO), we studied three different supercells, as shown in Fig.~\ref{fig:LSNO_struc}, with La as an example for the chemical formula $R_3$SrNi$_4$O$_8$. 
The \textit{flat} LSNO cell has three La ions and one Sr ion in each rare earth plane, resulting in a $2\times2\times1$ supercell compared to the pristine cell. 
In the \textit{stacked} LSNO cell, three LaNiO$_2$ layers and one SrNiO$_2$ layer are stacked along the $c$ axis resulting in a $1\times1\times4$ supercell. 
The \textit{mixed} LSNO cell is a mixture of the previous two cells resulting in a $\sqrt{2}\times\sqrt{2}\times2$ supercell.
In \textit{flat} LSNO and \textit{mixed} LSNO, the planar symmetry is broken and the Ni-O-Ni angle is expected to deviate from 180$^\circ$. 
The DFT+U relaxed LaNiO$_2$ and La$_{0.75}$Sr$_{0.25}$NiO$_2$ structural parameters are listed in Table \ref{tbl:LSNO_lattice}. 
The three different LSNO configurations have quite similar lattice constants. Their in-plane lattice constants decreased by 0.3 \AA$\:$ and the out-of-plane lattice constants increased by 0.4 \AA$\:$  compared to the LaNiO$_2$ structure. 
The \textit{flat} LSNO has the most bending of the Ni-O-Ni angle, 174.3$^\circ$. Compared with bulk LaNiO$_3$ (160.1$^\circ$), this deviation is much smaller, and its effect should thus be negligible. 
The total energies are close with differences smaller than 0.024 eV, showing that no specific Sr doping position is preferred in our calculations.  
The $a(c)-$axis of PSNO is 3.89(3.31)\AA\:, which is similar to PNO.

\begin{table}
    \centering
    \caption{Lattice constants and total energies of LaNiO$_2$ (LNO) and La$_{0.75}$Sr$_{0.25}$NiO$_2$ (LSNO). $a$, $b$, $c$ are the pseudo-cubic lattice parameters [\AA], $d_{Ni-O}$ is the Ni-O bond length [\AA] and $\phi$ is the Ni-O-Ni angle [$^o$].  $E$ is the total energy per formula unit [eV].} 
    \begin{tabular}{p{0.25\linewidth}|p{0.12\linewidth}p{0.12\linewidth}p{0.15\linewidth}p{0.12\linewidth}p{0.12\linewidth}}
    \hline\hline
        & $a,b$ [\AA] & $c$ [\AA] & $d_{Ni-O}$[\AA] & $\phi$ [$^o$]  & E [eV] \\ 
    \hline
    LNO       &3.93   &3.31   &1.97   &180    &-27.494\\
    LSNO--flat      &3.90   &3.35   &1.95   &174.3  &-26.279\\
    LSNO--stacked   &3.91   &3.35   &1.96   &180    &-26.301\\
    LSNO--mixed     &3.90   &3.35   &1.95   &178.4  &-26.302\\
    \hline
    PNO   &3.88   &3.33   & 1.94   & 180    & -26.695\\
    PSNO--mixed     & 3.89   & 3.31   & 1.94   & 178.4  &-25.632\\
    \hline
    \end{tabular}
\label{tbl:LSNO_lattice}
\end{table} 

\section{Doping effect for supercell vs VCA}
\label{appdx:doping-LSNO}

\begin{table}
    \centering
    \caption{Ni $3d$ orbital occupancy, $n_d$, and that for $d_{z^2}$ and $d_{x^2-y^2}$ orbitals in L(S)NO.} 
    \begin{tabular}{p{0.25\linewidth}|p{0.18\linewidth}p{0.18\linewidth}p{0.18\linewidth}}
    \hline\hline
        & $n_d$  & $n_{d_{x^2-y^2}}$  & $n_{d_{z^2}}$ \\ 
    \hline
    LNO      &8.62  &1.26  &1.63 \\
    LSNO--flat      &8.54  &1.11  &1.63 \\
    LSNO--stacked   &8.54  &1.11  &1.63 \\
    LSNO--mixed     &8.54  &1.11  &1.63 \\
    LNO (VCA)        &8.52  &1.11  &1.62 \\
    \hline
    \end{tabular}
\label{tbl:edmft_LSNO}
\end{table}

\begin{figure}
    \centering
    \includegraphics[width=1\linewidth]{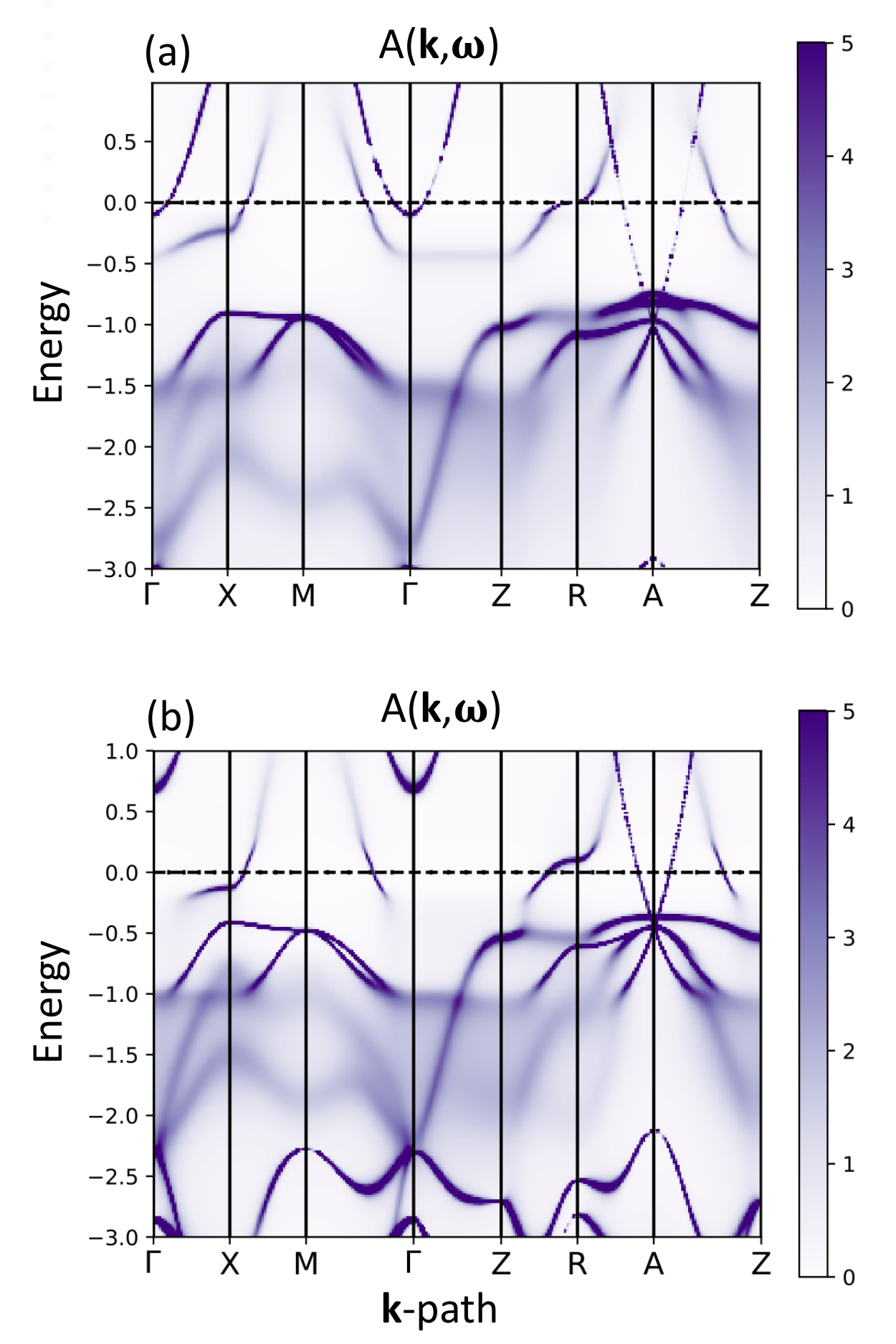}
\caption{The \textbf{k}-resolved spectral function, $A(\mathbf{k},\omega)$, of (a) LNO and (b) LNO with the 25\% hole doping (VCA) obtained from DFT+DMFT.}
\label{fig:BS_LNO_vca}
\end{figure}

Here, we study the hole-doping effect on LNO by comparing the supercell calculation and the virtual crystal approximation (VCA) results at the same 25\% doping.
In Table \ref{tbl:edmft_LSNO}, we list the occupancy of the Ni $e_g$ orbitals. 
Different LSNO supercells and LNO (VCA) exhibit similar occupancies since 
all $d_{x^2-y^2}$ occupancies are close to 1.11, decreased from the pristine LNO value of 1.26.
The $d_{z^2}$ orbitals for both L(S)NO and LNO (VCA) are not fully occupied, however, their occupancies do not change much upon hole doping, which is consistent with the PSNO result in the main text. The rest of the holes reside in La and O states.

Fig.~\ref{fig:BS_LNO_vca}(a) shows the \textbf{k}-resolved spectral function from DFT+DMFT for LaNiO$_2$. 
Among the $d$ orbitals, $d_{x^2-y^2}$ has the highest energy due to crystal field splitting and so is partially occupied near the Fermi energy. 
Other $d$ orbitals are mainly distributed below -1eV and also hybridized with the O 2$p$ states. 
La 5$d$ bands are mostly unoccupied while they are located below the Fermi level at the $\Gamma$ and $A$ points of the Brillouin zone with an energy as low as -1 eV.
These latter states are hybridized with Ni $d$ states of the relevant symmetries.
With hole doping calculated using the VCA method, we plot the \textbf{k}-resolved spectral function in Fig.~\ref{fig:BS_LNO_vca}(b).
Both the occupied Ni $d$ and O $p$ bands located below -1eV for LNO shift towards higher energy closer to the Fermi energy.
The La $5d$ bands also shift upwards, depopulating the $5d$ pocket at $\Gamma$ and shrinking the one at $A$. 
This trend is consistent with the orbital-resolved spectral function $A(\omega)$ results when comparing LNO to LSNO in the main text (Fig.\:\ref{fig:PNO_doping}).
This touching of La 5$d$ bands with the Ni 3$d$ bands at the $A$ point is preserved under hole doping and this differs from what has been presented for NdNiO$_2$ in previous work~\cite{NdNiO2_doping}. 

\section{Effects of the Hund's coupling $J$ and the crystal field basis}
\label{appdx:hundJ}

\begin{table}[h]
    \centering
    \caption{Orbital occupancies of PNO and PSNO compared for $J$=0.6eV vs 1.0eV from DFT+DMFT.}
    \begin{tabular}{p{0.18\linewidth}|p{0.17\linewidth}p{0.17\linewidth}p{0.17\linewidth}p{0.17\linewidth}}
    \hline
    Orbitals &  PNO ($J$=0.6eV)  &   PSNO ($J$=0.6eV)  &            PNO ($J$=1.0eV)  & PSNO ($J$=1.0eV)  \\
    \hline\hline
    Pr $4f$          &2.07  &2.06  &2.09 & 2.07  \\
    Ni $3d$  & 8.65 & 8.59 & 8.64 & 8.59 \\
    Ni $3d_{x^2-y^2}$  & 1.27 & 1.20 & 1.27 & 1.21  \\
    Ni $3d_{z^2}$  & 1.64     & 1.63 & 1.62 & 1.62 \\
    \hline
    \end{tabular}
    \label{tab:OCC_DC_Prf-2}
\end{table}

\begin{table}[h]
    \centering
    \caption{Statistical probabilities (in \%) of the many-body states of PNO and PSNO compared for $J$=0.6eV vs 1.0eV from DFT+DMFT.}
    \begin{tabular}{p{0.18\linewidth}|p{0.17\linewidth}p{0.16\linewidth}p{0.17\linewidth}p{0.17\linewidth}}
    \hline
    States &  PNO ($J$=0.6eV)  & PSNO ($J$=0.6eV) & PNO ($J$=1.0eV) & PSNO ($J$=1.0eV) \\
    \hline\hline
    Pr $f^2$          & 92.0  & 93.0  & 91.1 & 92.3 \\
    Pr $f^2_{J=4}$ & 91.6 & 92.5 & 90.7 & 91.8 \\
    Pr $f^3$          & 7.2  & 6.1  & 8.1 & 6.8 \\
    Ni $d^8_{S=0}$  & 12.2 & 17.0 & 7.9 & 9.6  \\
    Ni $d^8_{S=1}$  & 21.4  & 19.5 & 26.5 & 27.1 \\
    Ni $d^9$  & 52.8  & 50.7 & 51.9 & 50.3 \\
    Ni $d^{10}$  & 8.7  & 7.2 & 8.6 & 7.3 \\
    \hline
    \end{tabular}
    \label{tab:weight_DC_Prf-2}
\end{table}

It has been argued that the Hund's coupling $J$ of the Ni $3d$ orbitals can play an important role in nickelate superconductors as it can induce both a spin-state transition~\cite{PhysRevB.103.075123} and Hund's metal physics~\cite{PhysRevB.102.161118}.
Although all DFT+DMFT calculations in the main text used $J$=0.6eV for both Ni $3d$ and Pr $4f$ orbitals, here we compare this with DMFT results obtained using $J$=1eV for both Ni and Pr orbitals to investigate the effect of the Hund's coupling. 

\begin{figure}
    \includegraphics[width=1.0\linewidth]{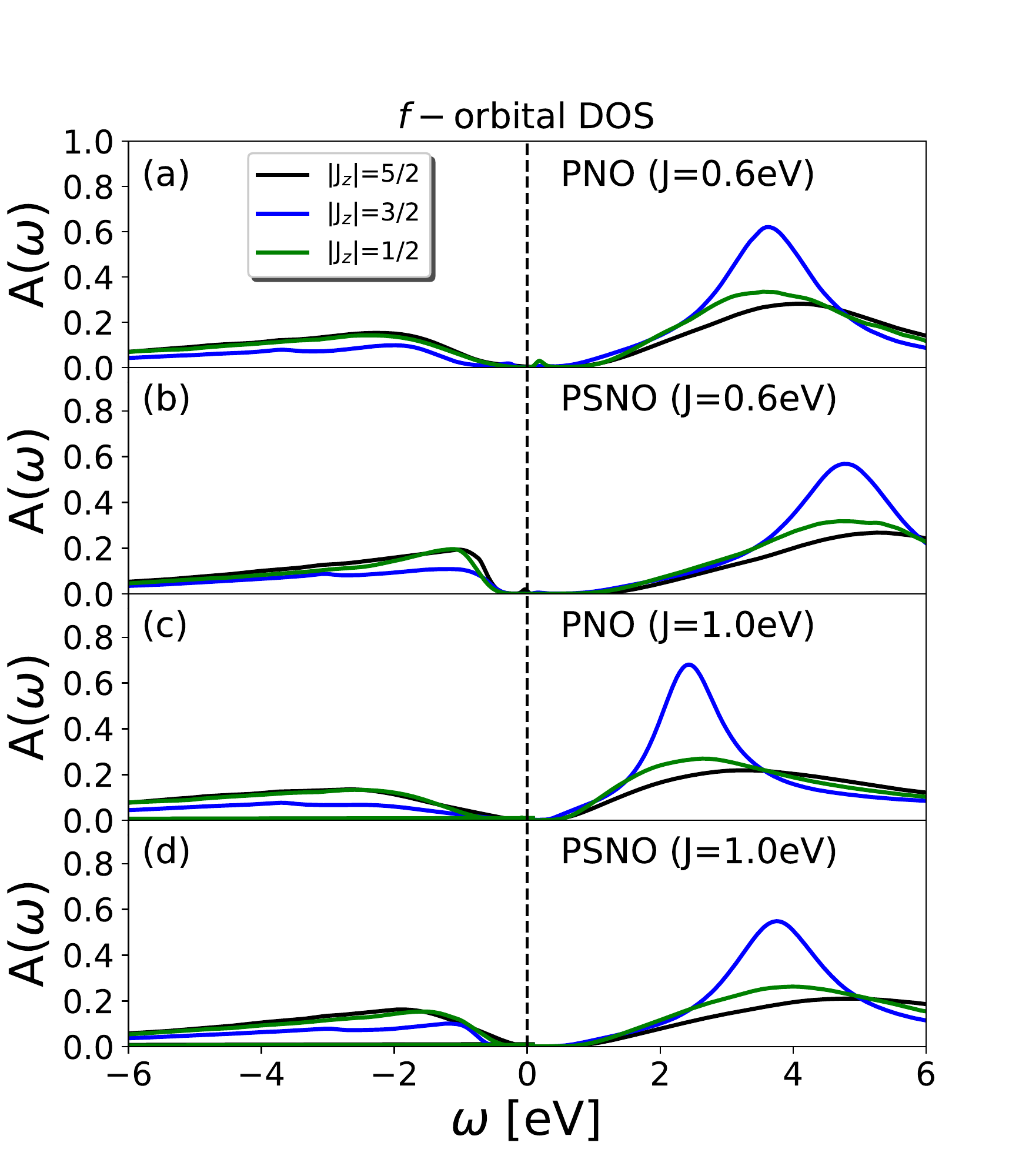}
\caption{The momentum-integrated spectral function $A(\omega)$ for Pr $4f$ $J=5/2$ orbitals in (a) PNO with $J$=0.6eV, (b) PSNO with $J$=0.6eV, (c) PNO with $J$=1.0eV, and (d) PSNO with $J$=1.0eV.}
\label{fig:PNO_f_J}
\end{figure}

\begin{figure}
    \includegraphics[width=1.0\linewidth]{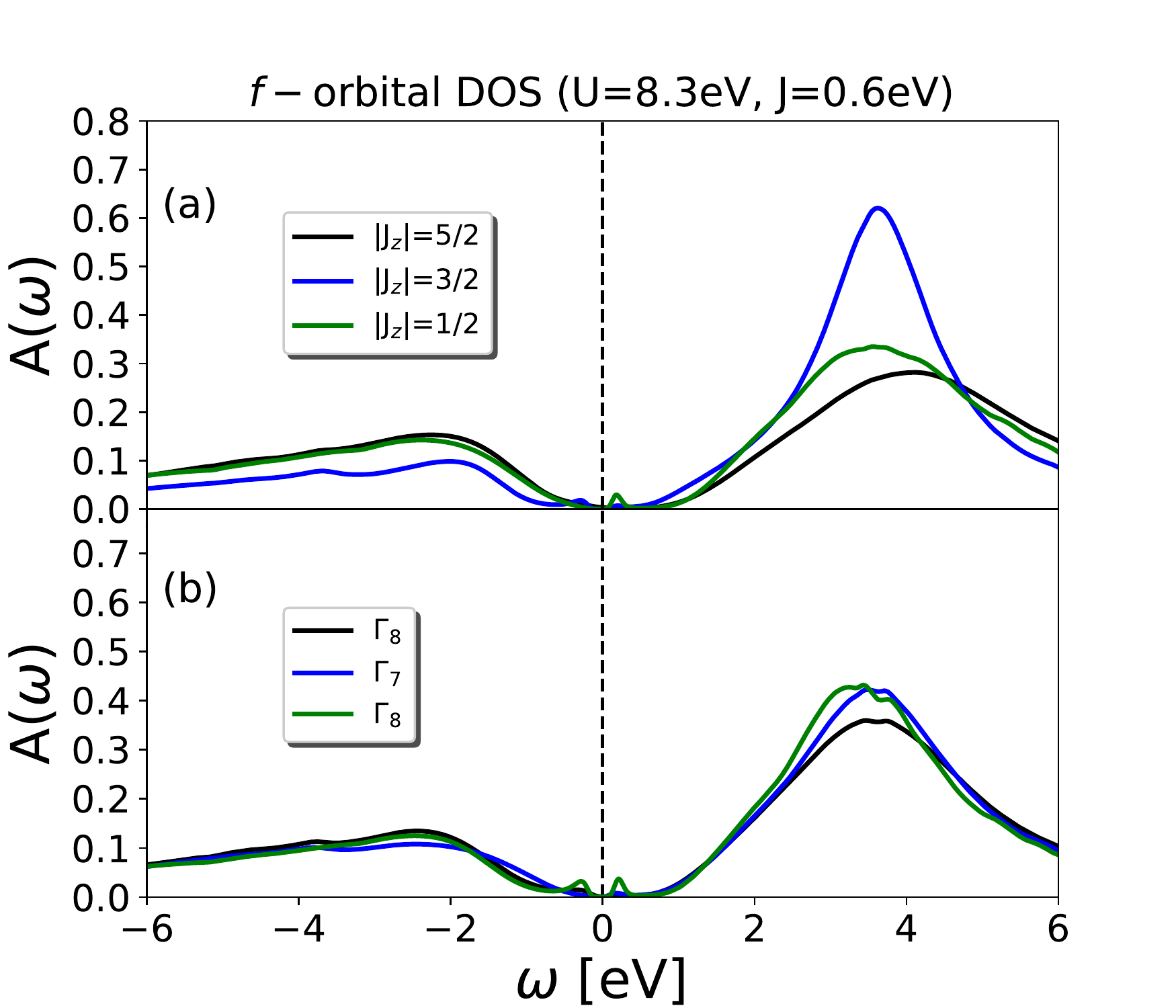}
\caption{The momentum-integrated spectral function $A(\omega)$ for Pr $4f$ $J=5/2$ orbitals computed using DFT+DMFT in (a) a $J_z$ basis and (b) a cubic crystal-field basis.  For the latter, the $\Gamma_8$ result in green is the same as $|J_z|=1/2$.}
\label{fig:PNO_f_c}
\end{figure}

\begin{table}[h]
    \centering
    \caption{The mass renormalization of PNO and PSNO compared for $J$=0.6eV vs 1.0eV from DFT+DMFT.}
    \begin{tabular}{p{0.18\linewidth}|p{0.16\linewidth}p{0.16\linewidth}p{0.16\linewidth}p{0.16\linewidth}}
    \hline
    Orbitals &   PNO ($J$=0.6eV)  & PSNO ($J$=0.6eV) & PNO ($J$=1.0eV) & PSNO ($J$=1.0eV)  \\
    \hline\hline
    Ni $3d_{x^2-y^2}$ & 1.94 & 1.68 & 2.27 & 1.92           \\
    \hline
    Ni $3d_{z^2}$ & 1.26 & 1.25 & 1.28 & 1.29          \\
    \hline
    \end{tabular}
    \label{tbl:mass-2}
\end{table}

The change of the Hund's coupling $J$ from 0.6eV to 1.0eV does not affect the orbital occupancy of Pr $4f$ and Ni $3d$ orbitals for both the PNO and PSNO cases (Table \ref{tab:OCC_DC_Prf-2}). However, as expected, the probability for the Ni $d^8_{S=1}$ high spin state is noticeably increased for $J$=1.0eV in both PNO and PSNO (Table \ref{tab:weight_DC_Prf-2}). As a result, the mass renormalization of the Ni $d_{x^2-y^2}$ orbital is also increased by $\sim$17\% for $J$=1.0eV, though that of the Ni $d_{z^2}$ orbitals is not much affected (Table \ref{tbl:mass-2}). More importantly, the Pr $4f$ spectral functions for $J$=1.0eV in Fig.~\ref{fig:PNO_f_J} shows that the spectral gap at the Fermi energy is still retained in both the PNO and PSNO cases.  That is, the Kondo resonance observed in the GW+DMFT study~\cite{Kang22} is absent in our calculations regardless of the different $J$ values we used.
This is consistent with the dominance of the $f^2$ $J=4$ configuration in Table \ref{tab:weight_DC_Prf-2}.

Finally, we present in Fig.~\ref{fig:PNO_f_c} a comparison of $A(\omega)$ for the $4f$ states in a $J_z$ basis and a cubic crystal-field basis (approximate for tetragonal PrNiO$_2$).  The latter shows less orbital differentiation than the former.  This is a reflection of the weak influence of the crystal-field splitting relative to that due to the multiplets.  We also note that the relevant $Y_{3\pm2}$ harmonics for hybridization of $4f$ and oxygen $2p$ are more dominant for $|J_z|= 3/2$ (71.4\%) than for $\Gamma_7$ (61.9\%).

\end{document}